\documentclass[fleqn,usenatbib]{mnras}

\usepackage{newtxtext,newtxmath}

\usepackage[T1]{fontenc}

\DeclareRobustCommand{\VAN}[3]{#2}
\let\VANthebibliography\thebibliography
\def\thebibliography{\DeclareRobustCommand{\VAN}[3]{##3}\VANthebibliography}

\usepackage{graphicx}	
\usepackage{amsmath}
\usepackage{ulem}

\newcommand{\fermi}{{\it Fermi}-LAT}
\newcommand{\gray}{$\gamma$-ray}
\newcommand{\grays}{$\gamma$-rays}
\newcommand{\source}{BL Lac}
\newcommand{\gsim}{{\lower.5ex\hbox{$\; \buildrel > \over \sim \;$}}}

\title[Multi-wavelength emission from BL Lac]{A thirteen-year-long broadband view of BL Lac}

\author[N. Sahakyan \& P. Giommi]{
N. Sahakyan,$^{1,2, 3}$ \thanks{E-mail: narek@icra.it}
P. Giommi$^{2,4,5,6}$
\\
$^{1}$ICRANet-Armenia, Marshall Baghramian Avenue 24a, Yerevan 0019, Armenia\\
$^{2}$ICRANet, P.zza della Repubblica 10, 65122 Pescara, Italy\\
$^{3}$ ICRA, Dipartimento di Fisica, Sapienza Universita` di Roma, P.le Aldo Moro 5, 00185 Rome, Italy\\
$^{4}$ Associated to Italian Space Agency, ASI, via del Politecnico snc, 00133 Roma, Italy\\
$^{5}$Institute for Advanced Study, Technische Universit{\"a}t M{\"u}nchen, Lichtenbergstrasse 2a, D-85748 Garching bei M\"unchen, Germany\\
$^{6}$ Center for Astro, Particle and Planetary Physics (CAP3), New York University Abu Dhabi, PO Box 129188 Abu Dhabi, United Arab Emirates;
}

\date{Accepted XXX. Received YYY; in original form ZZZ}
\pubyear{2015}

\begin{document}
\label{firstpage}
\pagerange{\pageref{firstpage}--\pageref{lastpage}}
\maketitle

\begin{abstract}
We present the results of an extensive analysis of the optical, ultraviolet, X-ray and $\gamma$-ray data collected from the observations of the BL Lac objects prototype BL Lacertae carried out over a period of nearly 13 years, between August 2008 and March 2021.   The source is characterized by strongly variable emission at all frequencies, often accompanied by spectral changes. In the $\gamma$-ray band  several prominent flares have been detected, the largest one reaching the flux of $F_{\rm \gamma}(>196.7\: {\rm MeV})=(4.39\pm1.01)\times10^{-6}\:{\rm photon\:cm^{-2}\:s^{-1}}$. The X-ray spectral variability of the source during the brightest flare on MJD 59128.18 (06 October 2020) was characterized by a softer-when-brighter trend due to a shift of the synchrotron peak to $\sim 10^{16}$ Hz, well into the HBL domain. The widely changing multiwavelength emission of BL Lacertae was systematically investigated by fitting leptonic models that include synchrotron self-Compton and external Compton components to 511 high-quality and quasi-simultaneous broad-band spectral energy distributions (SEDs). The majority of selected SEDs can be adequately fitted within a one-zone model with reasonable parameters. Only 46 SEDs with soft and bright X-ray spectra and when the source was observed in very high energy $\gamma$-ray bands can be explained in a two-zone leptonic scenario. The HBL behaviour observed during the brightest X-ray flare is interpreted as due to the emergence of synchrotron emission from freshly accelerated particles in a second emission zone located beyond the broad line region.
\end{abstract}

\section{Introduction}
Radio-loud Active Galactic Nuclei (AGNs) are characterized by two-sided narrow relativistic jets that originate from the central supermassive black hole. Blazars are the subclass of radio loud AGNs in which one of the jets happens to make a small angle ($<10^{\circ}$) to the line of sight of the observer \citep{1995PASP..107..803U}. These jets transport a large amount of power in the form of particles, radiation and magnetic field and are strong sources of non-thermal emission. Due to the small viewing angle and the relativistic motion the emission in blazars is strongly Doppler boosted, a special situation that makes these sources detectable up to large redshifts \citep[e.g.,][]{2017ApJ...837L...5A, 2020MNRAS.498.2594S} and is responsible for the observed extreme properties that characterizes them, like superluminal motion and rapid variability across the electromagnetic spectrum. Historically blazars are classified as BL Lacertae objects (BL Lacs), which exhibit an optical spectrum that is completely featureless or at most shows very weak emission lines (equivalent width ${\rm EW}\leq5{\AA}$), and as flat spectrum radio quasars (FSRQs) when the emission lines are stronger and quasar-like \citep{1995PASP..107..803U}. Blazars are generally assumed to be persistent sources, however a case of a transient blazar, 4FGL J1544.3-0649, was recently observed. 
This object remained below the sensitivity limits of X-ray and \gray{} instruments until May 2017 when it raised above detactability and for a few months it became one of the brightest X-ray blazars \citep{2021MNRAS.502..836S}. If this was not an isolated case, but rather a common phenomenon, it could have an impact on the real abundance and on our current understanding of blazars.\\
\indent The broadband SED of blazars, in a ${\rm log(\nu F\nu)}$ vs. ${\rm log(\nu)}$ representation, shows two prominent broad components, one (low-energy component) peaking form far infrared frequencies to X-ray energies and another (high energy component) peaking at MeV/GeV energies. The peak of the low-energy component ($\nu_{\rm s}$) is used to further classify blazars as high synchrotron peaked BL Lacs (HBL when $\nu_{\rm s}>10^{15}$ Hz), intermediate synchrotron peaked BL Lacs (IBL when $10^{14}<\nu_{\rm s}<10^{15}$ Hz), or low synchrotron peaked BL Lacs (LBL when $\nu_{\rm s}<10^{14}$ Hz) objects \citep{Padovani1995,Abdo_2010}. Sometimes the synchrotron peak can reach energies as high as $\sim$1 keV, ($\sim 2\times 10^{17}$ Hz) or beyond, showing what is considered to be extreme behaviour, even for these highly peculiar sources \citep[e.g.][]{sedentary,2001A&A...371..512C,Biteau2000}. Such a high synchrotron peak 
was first observed during a flare of Mkn 501 \citep{1998ApJ...492L..17P}, and subsequently in many other objects \citep[e.g., ][]{2018MNRAS.477.4257C, 2020MNRAS.496.5518S}. Independently of the location of the peak, the low-energy part of the SED is generally interpreted as synchrotron emission from the relativistic electrons in the jet. A proton synchrotron origin of the high energy end of this component during X-ray flares has also been considered \citep{2021ApJ...906..131M,2021arXiv210714632S}.
The nature of the high energy (HE; $>100$ MeV) component is instead still under debate. Within one-zone leptonic scenarios, the second component originates from inverse Compton scattering of the synchrotron photons (SSC) by the electron population producing the low-energy component \citep{ghisellini, bloom, maraschi}. Depending on the location of the emission region, the photons external to the jet (e.g., photons from the disc, or those reprocessed from the broad-line region or those from the infrared torus) can up-scatter, producing the second component \citep[external inverse Compton (EIC);][]{blazejowski,ghiselini09, sikora}. On the other hand, the HE component can be also produced from the interaction of relativistic protons either from their synchrotron emission \citep{2001APh....15..121M} or from the secondary particles from pion decay \citep{1993A&A...269...67M, 1989A&A...221..211M, 2001APh....15..121M, mucke2, 2013ApJ...768...54B}. Recently, after associating TXS 0506+056 with the IceCube-170922A neutrino event \citep{2018Sci...361..147I, 2018Sci...361.1378I, 2018MNRAS.480..192P} the lepto-hadronic scenarios, when both electrons and protons contribute to the HE emission, have become more attractive. These models also predict very high energy (VHE; $>100$ GeV) neutrinos observable by the IceCube detector \citep{2018ApJ...863L..10A,2018ApJ...864...84K, 2018ApJ...865..124M, 2018MNRAS.480..192P, 2018ApJ...866..109S, 2019MNRAS.484.2067R,2019MNRAS.483L..12C, 2019A&A...622A.144S, 2019NatAs...3...88G, 2022MNRAS.509.2102G}.

Blazars, being powerful sources of strongly variable non-thermal emission, are often targets of multiwavelength observations. The resulting data have been accumulating over time enriching the archives with very valuable information that can be used for detailed energy and time-domain investigations of the origin of their emission. BL Lacertae (\source) is one of these frequently studied blazars; at $z = 0.069$ it is a prototype of the BL LAC subclass of blazars. 
\source\ is usually classified as an LBL \citep{2018A&A...620A.185N}, but is sometimes listed as an IBL \citep{2011ApJ...743..171A}. \source\ is well known for its prominent variability in a wide energy range, especially in the optical \citep{2010A&A...510A..93L, 2015MNRAS.450..541A} and radio bands \citep{2016ApJ...816...53W}. \source\ has been a target of many multiwavelength campaigns ranging from the radio to the HE or VHE \gray{} bands \citep{2008Natur.452..966M, 2009A&A...507..769R, 2013MNRAS.436.1530R, 2019A&A...623A.175M, 2020ApJ...900..137W} which resulted in a deep understanding of its properties in different bands. For example, in the X-ray band, BeppoSAX observations in June 1999 showed that the 0.3-2 keV flux of \source\ doubled in $\sim20$ min and the spectrum was concave with a very hard component above 5-6 keV \citep{2002A&A...383..763R}. In the \gray\ band, the EGRET observations in 1995 showed an average \gray\ flux above 100 MeV of $(40\pm12)\times10^{-8}\:{\rm photon\:cm^{-2}\:s^{-1}}$ \citep{1997ApJ...480..562C} which increased up to $(171\pm42)\times10^{-8}\:{\rm photon\:cm^{-2}\:s^{-1}}$ during the flare in 1997 \citep{1997ApJ...490L.145B}. Afterwards, the observations by the Large Area Telescope (LAT) on board the Fermi Gamma-ray Space Telescope (\fermi)  showed that during flaring periods the average \gray\ flux above 100 MeV can reach above $10^{-6}\:{\rm photon\:cm^{-2}\:s^{-1}}$ \citep[e.g., see][]{2012ATel.4028....1C, 2020ATel13933....1C,2020ATel14072....1M,2020ATel13964....1O,2021ATel14330....1C}. VHE $\gamma$-rays above 1 TeV from \source\ were initially reported by the Crimean Observatory in 1998 \citep{2001ARep...45..249N} and later, in 2005, the MAGIC telescope discovered a VHE \gray\ signal with an integral flux of 3\% of the Crab Nebula flux above 200 GeV \citep{2007ApJ...666L..17A}. The source is flaring also in the VHE \gray\ band; for example, on June 28 2011, a very rapid TeV \gray\ flare was detected by VERITAS when the integral flux above 200 GeV reached roughly 125\% of the Crab Nebula flux \citep{2013ApJ...762...92A}, or 
on June 15 2015 MAGIC detected a flare with a maximum flux of $(1.5 \pm 0.3) \times 10^{-10} {\rm photons\: cm^{-2}\: s^{-1}}$ and halving time of $26 \pm 8$ min \citep{2019A&A...623A.175M}.\\
\indent \source\ shows a peculiar behavior both in terms of its classification and interpretation of the observed broadband SED. First, the observation of $H\alpha$ and $H\beta$ lines ($\sim10^{41}\:{\rm erg\: s^{-1}}$) \citep{1996MNRAS.281..737C,2010A&A...516A..59C} in different periods is quite unusual for this type of blazars. This might indeed indicate a presence of a broad-line region structure. On the other hand, the single-zone SSC models, usually successful for explaining the TeV BL Lac spectrum, have a difficulty in reproducing the variability of this source in different bands and taking into account the emission in all the bands. When the spectrum extends to the VHE \gray\ band or when a large Compton dominance is observed, the SED of \source\ can be modeled only by considering an EIC component added to SSC or by using two-zone models \citep[e.g.,][]{1997ApJ...490L.145B, 1999ApJ...521..145M, 2000AJ....119..469B, 2011ApJ...730..101A, 2019A&A...623A.175M}. This illustrates that different models/components are contributing in the overall complex broadband spectrum of \source.\\
\indent Over the past decade, \source\ was constantly monitored in the HE \gray\ band by \fermi\ \citep{2020ApJ...892..105A} and AGILE \citep{2aglcat} and frequently observed in the optical/UV and X-ray bands by \textit{Neil Gehrels Swift Observatory} \citep{2004ApJ...611.1005G}, (hereafter \textit{Swift}). Together with the observations of other instruments (\textit{NuSTAR}, MAGIC, VERITAS, etc.) this resulted in the accumulation of an extremely rich multi-frequency data set mapping both emission components. The available data can be combined to build the broadband SED of \source\ in many different periods with (quasi) contemporaneous data. The theoretical interpretation of these SEDs can help understanding the physical processes that dominate in different periods. For example, a similar study of the broadband emission of 3C 454.3 allowed us to estimate the main parameters describing the jet and emitting electrons as well as to investigate their evolution in time \citep{2021MNRAS.504.5074S}. Moreover, \source\ was in active flaring states from optical to \gray\ bands in October 2020 and January 2021 \citep[e.g., ][]{2021ATel14318....1M, 2021ATel14328....1M, 2021ATel14330....1C, 2021ATel14342....1D, 2021ATel14334....1H, 2021ATel14350....1D} when the brightest \gray\ flare from this source was also observed \citep{2020ATel14072....1M}; on October 6 2020, the daily averaged \gray\ flux of \source\ was $(5\pm1) \times 10^{-6} \: {\rm photons\:cm^{-2}\:s^{-1}}$. The available multiwavelength data and the extraordinary flaring activity of \source\ in 2020/2021 motivated us to have a new look on the origin of the broadband emission from it.\\
\indent In this paper, analyzing the data observed by \fermi, \textit{Swift} X-ray Telescope (XRT) and Ultraviolet and Optical Telescope (UVOT) accumulated in the previous thirteen years, we perform an intense broadband study of \source. The paper is organized as follows. The \fermi\ and \textit{Swift} data collected for the analysis and its reduction methods are described in Section \ref{datanal}. The spectral changes in different bands and the broadband SED modeling is discussed in Section \ref{modeling}. The discussion is presented in Section \ref{discussion} and the summary in Section \ref{concl}.
\section{\fermi\ Observations and Data Analysis}\label{datanal}

\indent Since August 2008, \source\ was constantly observed by \fermi\  providing unprecedented information on its emission in the \gray\ band.
\fermi\ is a pair conversion telescope sensitive to \grays\ in the energy range from 100 MeV to 500 GeV. By default, it operates in all sky scanning mode, mapping the entire \gray\ sky every three hours. Further details on \fermi\ are given in \citet{2009ApJ...697.1071A}.

For the current study, publicly available data accumulated between 04 August, 2008 and 01 March, 2021 are used (MET 239557417 - 636249605). The data have been analysed by using Fermi ScienceTools  version 1.2.1. The Pass8 “Source” class events with a higher probability of being photons ({\it evclass = 128}, {\it evtype=3}) in the energy range from 100 MeV to 500 GeV were analyzed using P8R3\_ SOURCE\_ V3 instrument response function. The events were downloaded from a region of interest (ROI) defined as a circular region with $12^{\circ}$ around the \gray\ position of \source. The events are binned within a $16.9^{\circ} \times 16.9^{\circ}$ square region into pixels of $0.1^{\circ}\times0.1^{\circ}$ and into 37 equal logarithmically spaced energy bins. The model was created using the \fermi\ fourth source catalog Data Release 2 \citep[4FGL-DR2; ][]{2020ApJ...892..105A} where all sources within $17^{\circ}$ around the target as well as the Galactic (gll\_ iem\_ v07) and the isotropic (iso\_ P8R3\_ SOURCE \_ V3\_ v1) diffuse emission components are included. The spectral parameters of the background sources falling between $12^{\circ}$ and $12^{\circ}$+$5^{\circ}$ were fixed to their catalog values, while the parameters of the other sources and background models were left free. Binned likelihood analysis was applied with {\it gtlike} tool to find the best matches between spectral models and the data. The source variability was investigated by dividing the entire period to three-day bins. During these short periods the source spectrum was modeled using a power-law function, and the  photon flux and index were estimated by applying unbinned likelihood analysis with the appropriate quality cuts mentioned above. The light curves were computed by fixing the spectral indices of all sources (except \source) and the normalization of both the Galactic and isotropic components to the best-fit values obtained for the whole time period and then by allowing them to vary. 
In all cases the light-curves are fully consistent with each other and with the one available in the \fermi\ light curve repository \footnote{\url{https://fermi.gsfc.nasa.gov/ssc/data/access/lat/LightCurveRepository/index.html}}. 
In addition to the three-day binned light curve, an adaptively binned light curve was generated by adjusting the time bin widths so as to attain 20\% uncertainty in the flux estimation above an optimal energy \citep[see ][for details]{2012A&A...544A...6L}. This light curve with unequal time bins has been proven to be particularly efficient for the identification of flaring states \citep[e.g., see][]{2018ApJ...863..114G, 2018A&A...614A...6S, 2017A&A...608A..37Z, 2017ApJ...848..111B, 2016ApJ...830..162B, 2013A&A...557A..71R}.\\ 
\begin{figure*}
	\includegraphics[width=0.9\textwidth]{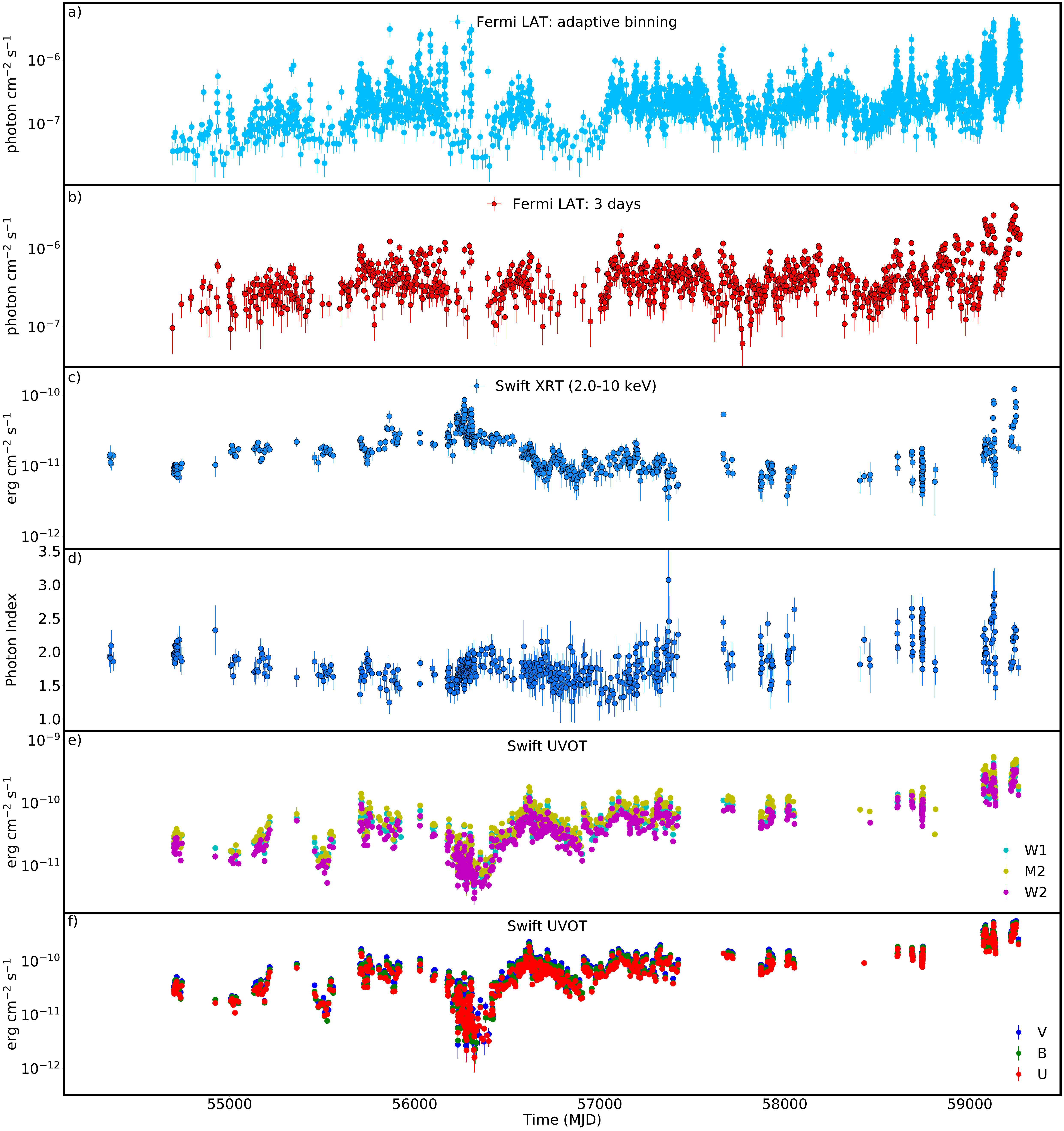}
    \caption{The multiwavelength light curve of \source\ between August 04, 2008 and March 01, 2020. \textit{a)} Adaptively binned \gray\ light curve ($>196.7$ MeV); \textit{b)} 3-day binned \gray\ light curve ($>100$ MeV); \textit{c)} \textit{Swift} XRT light curve in the 2.0-10 keV range; \textit{d)} 0.3-10.0 keV X-ray photon index; \textit{e, f)} host galaxy corrected flux in W1, M2, W2, V, B, and U filters.}
    \label{lightcurve_all}
\end{figure*}
The adaptively binned (E $>196.7$ MeV) and three-day (E $>100$ MeV) \gray\ light curves are shown in Fig. \ref{lightcurve_all} panels a) and b) respectively. The time-averaged \gray\ flux of \source\ above 100 MeV is $(3.71\pm0.05)\times10^{-7}\:{\rm photon\:cm^{-2}\:s^{-1}}$. Both light curves show the complex behaviour of \source\ in the \gray\ band; the mean \gray\ flux of the source is $4.46\times10^{-7}\:{\rm photon\:cm^{-2}\:s^{-1}}$ 
which increases up to $(4.39\pm1.01)\times10^{-6}\:{\rm photon\:cm^{-2}\:s^{-1}}$ 
(above $196.7$ MeV) observed on MJD 59231.34 (17 January 2021). Using the adaptively binned light curve in the considered thirteen years the source flux was above  $10^{-6}\:{\rm photon\:cm^{-2}\:s^{-1}}$ in total for $41.5$ days. The photon index variation in time is investigated using a 3-day binned light curve. The photon index is mostly soft with a mean value of $\Gamma_{\rm mean}=2.15$ but occasionally it hardened to  $\Gamma <2.0$. The hardest indexes of $1.48\pm0.22$ and $1.61\pm0.17$ were observed on MJD 57771.16 (18 January 2017) and 55782.16 (09 August 2011), respectively.\\
\indent The time evolution of the \gray\ emission is also investigated by generating the SED at different times. When the SEDs are constructed for short periods (e.g., three-day bins or for the time intervals identified in the adaptive bins lightcurve) the spectrum can be measured only up to the moderate energies, not enough for a detailed study. Therefore, the Bayesian block algorithm \citep{2013ApJ...764..167S} is used to divide the \gray\ light curve into optimal intervals which are represented by an approximately constant flux. By applying this algorithm, the points where the flux changes from one state to another will be identified, providing the \gray\ spectra of the source in different states. The Bayesian block algorithm applied to the adaptively binned light curve divides the entire period into 218 intervals with a similar flux level. The shortest period is $5.81$ hours during a flare while in the low emission state the longest period is $278.74$ days. The spectral analysis is applied by limiting the time for each interval selected based on the Bayesian block. During the analysis, the spectrum of \source\ is assumed to be a power-law with spectral index and normalization left as free parameters. The best matches between the spectral models and events are obtained with an unbinned likelihood analysis implemented in {\it gtlike}. Depending on source intensity the spectrum of \source\ is obtained by separately running the analysis for 4 or 7 energy bands of equal width in log scale.
\begin{figure*}
	\includegraphics[width=0.48\textwidth]{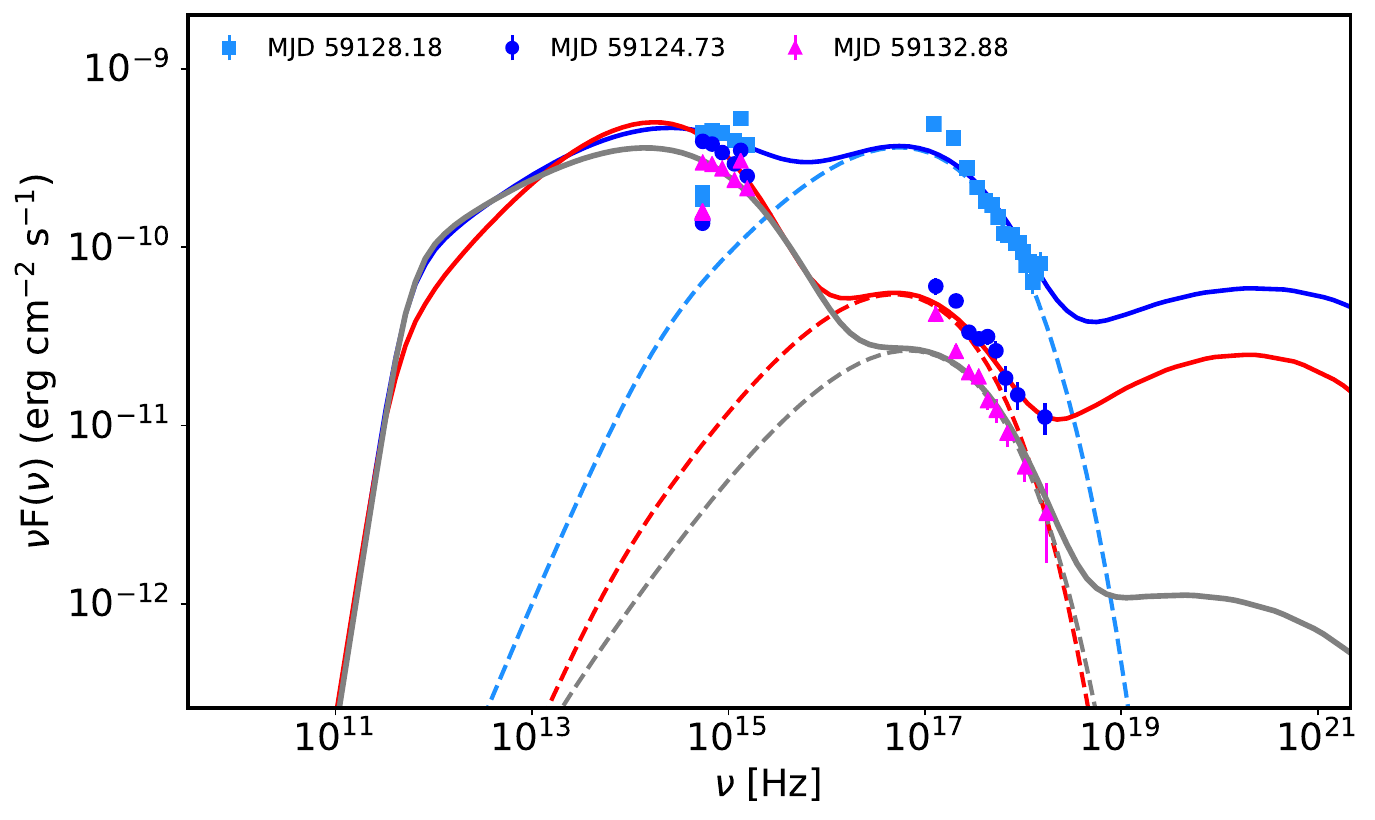}
	\includegraphics[width=0.48\textwidth]{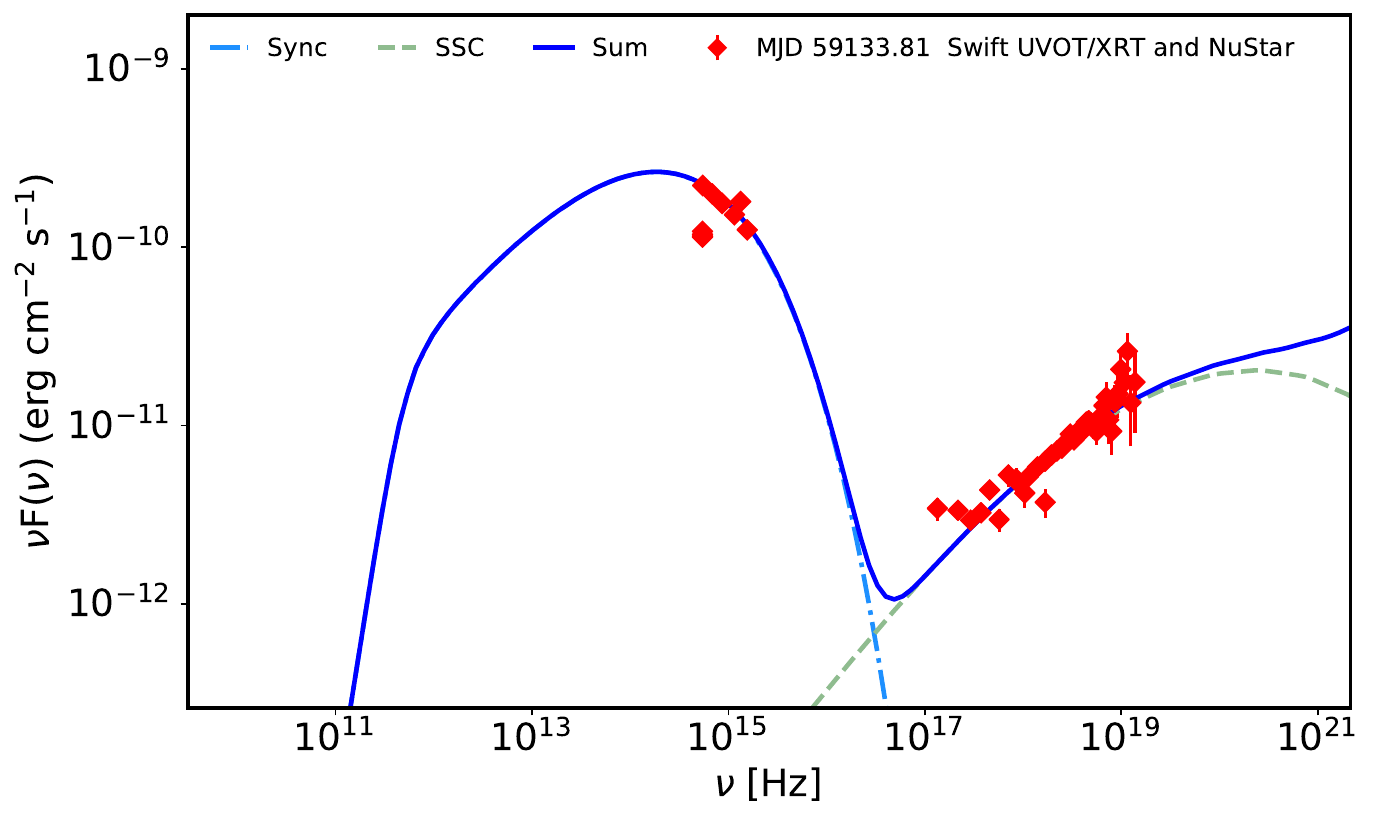}
    \caption{The SED of \source\ in different periods. {\it Left panel:} The periods with a soft X-ray spectrum as modeled within the two-zone leptonic scenario. The dashed lines show the synchrotron emission from the second region and the solid lines are the sum of the contribution from both regions.{\it Right panel:} The usual hard X-ray spectrum (\textit{Swift} XRT and \textit{NuSTAR}) observed on MJD 59133.88 (11 October 2020).}
    \label{only_syn}
\end{figure*}
\subsection{\textit{Swift} XRT}
During the considered period, the \textit{Swift} satellite observed \source\ 610 times with single exposures ranging from 1.13 to 16.46 ks. All the data were downloaded and processed using {\it Swift\_xrtproc} automatic tool for XRT data analysis developed within the Open Universe Initiative \citep{2021arXiv210807255G}. This tool automatically downloads the raw data and processes it using the XRTPIPELINE task adopting standard parameters and filtering criteria. For each observation, it extracts the source events from a circle with a radius of 20 pixels centered at the position of the source, while the  background counts are taken from an annular ring centered at the source. The tool applies also pile-up correction when the source count rate is above $0.5\:{\rm counts\: s^{-1}}$.  Then it loads the ungrouped data in XSPEC (version 12.11) for spectral fitting using Cash statistics \citep{1979ApJ...228..939C}, modeling the source spectrum as a power-law and a log-parabola model with the Galactic absorption column density fixed to  $2.7\times10^{21}\:{\rm cm^{-2}}$ \citep[e.g.,][]{1999ApJ...521..145M,2020ApJ...900..137W,2021MNRAS.tmp.2396D}.

The 2-10 keV X-ray flux variation is shown in Fig. \ref{lightcurve_all} panel c). The baseline flux is around $2\times10^{-11}\:{\rm erg\:cm^{-2}\:s^{-1}}$ although small amplitude changes are visible in different observations. In three periods, MJD 56300 (08 January 2013), MLD 59140 (18 October 2020) and MJD 59235 (21 January 2021), the flux substantially increased reaching the maximum of $(1.41\pm0.06)\times10^{-10}\:{\rm erg\:cm^{-2}\:s^{-1}}$ on MJD 59128.18 (06 October 2020). This is the historical highest flux of \source\ in the soft X-ray band. 
 
\begin{figure*}
	\includegraphics[width=0.48\textwidth]{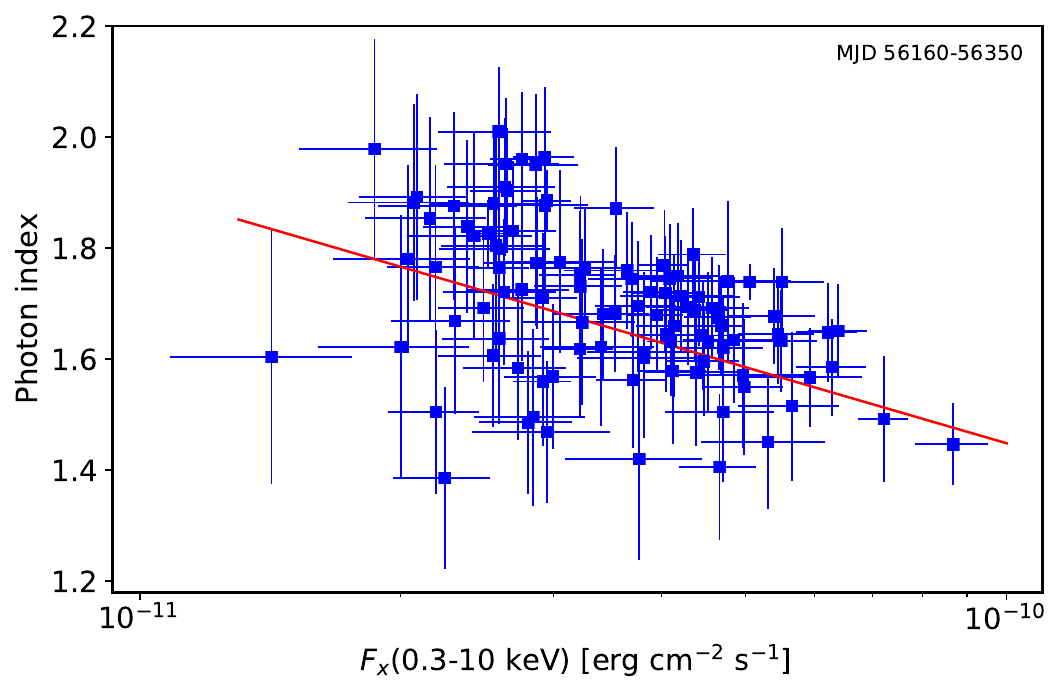}
	\includegraphics[width=0.48\textwidth]{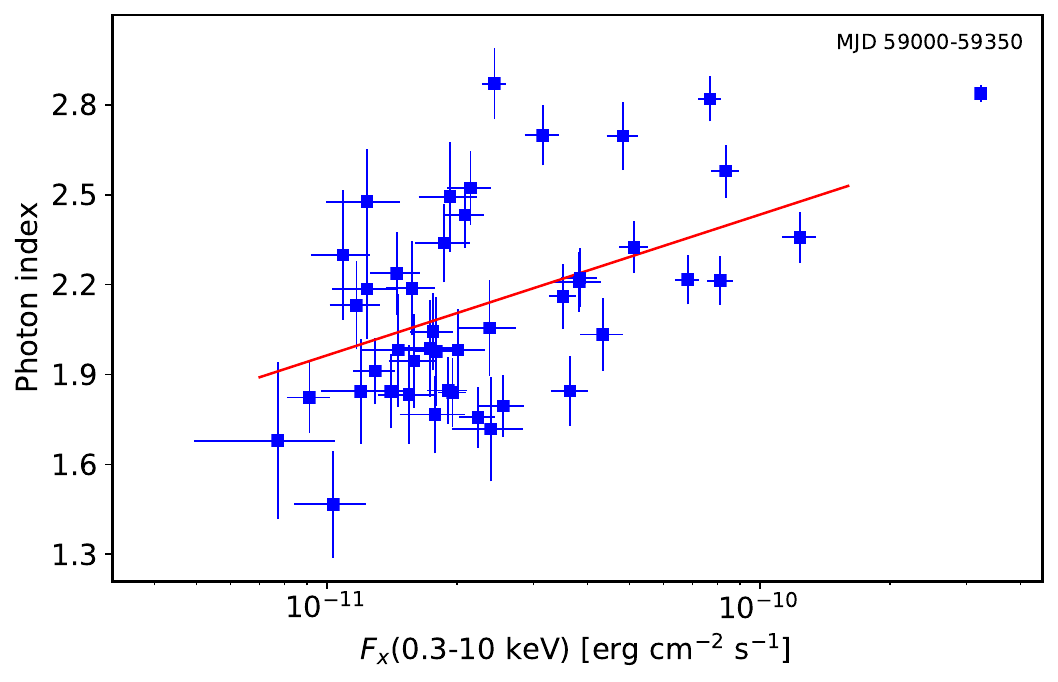}
    \caption{\source\ X-ray photon index versus the flux during two major X-ray flares. The correlation trend is shown with a red line.
    }
    \label{X:infl}
\end{figure*}

The X-ray photon index in different observations is shown in Fig. \ref{lightcurve_all} panel d). Most of the time, the photon index is hard ($\leq2.0$) implying that the X-ray emission is due to the rising part of the HE component in the SED of \source. However, the photon index undergoes interesting modifications reaching $2.0$ which corresponds to a flat distribution in $\nu f \nu$ vs  $\nu$ representation. For example, such tendency can be noticed after the X-ray flare around MJD 56300 (08 January 2013).
In the considered periods, also a significant softening of the photon index is observed; e.g., in 36 observations the X-ray photon index is $>2.3$ (considering only the observations when the number of counts was $>100$) which is unusual for \source\ and more typical of HBL blazars. Examples of optical/UV and X-ray spectrum of \source\ during such changes are shown in Fig \ref{only_syn}. The X-ray component started to soften starting from MJD 59113.16 (21 September 2020) when an index of $\Gamma_{\rm X}=2.43\pm0.11$ was observed. Then, the photon index softens to $\Gamma_{\rm X}=2.84\pm0.03$ on MJD 59128.18 (06 October 2020) during the brightest X-ray emission state (light blue squares in Fig. \ref{only_syn}).
In this period the optical/UV flux increased substantially as well, showing that the low-energy component now extends to the X-ray band. 
Such soft X-ray emission with $\Gamma_{\rm X}=2.82\pm0.07$ and a  flux of $F_{\rm X (0.3-10\:keV)}=(7.68\pm0.47)\times10^{-11}\:{\rm erg\:cm^{-2}\:s^{-1}}$ (red circle in Fig.\ref{only_syn}) was also observed on MJD 59128.91 (06 October 2020). In the next two observations (MJD 59129.90 (07 October 2020) and 59131.83 [09 October 2020]), the X-ray flux was constantly decreasing and the photon index was $\Gamma_{\rm X}=2.52-2.70$. The softest photon index of $\Gamma_{\rm X}=2.87\pm0.11$ was observed on MJD 59132.88 (10 October 2020; magenta triangles in Fig. \ref{only_syn}) when the source flux was $F_{\rm X (0.3-10\:keV)}=(2.44\pm0.16)\times10^{-11}\:{\rm erg\:cm^{-2}\:s^{-1}}$. 
However, this component fades in the next observations (e.g., on MJD 59133.81 [11 October 2020]) and in the X-ray band the usual HE component is observed.
There were additional periods when softening in the X-ray band was observed ($\Gamma_{\rm X}\geq2.5$); for example, on MJD 58685.98 (21 July 2019) and 58686.90 (22 July 2019) and between MJD 58740.42-58741.41 (14-15 September 2019) the X-ray photon index was $\Gamma_{\rm X}=2.51-2.64$ with a flux between $F_{\rm X (0.3-10\:keV)}=(8.36-17.64)\times10^{-12}\:{\rm erg\:cm^{-2}\:s^{-1}}$.\\
\indent The X-ray flux evolution was further investigated by comparing it with the photon index in different states. When considering the entire observational period with diverse X-ray emission properties, any trend (if present) will be smoothed out. For this reason, the X-ray photon index versus the flux was investigated by selecting the periods around two major flares visible in Fig. \ref{lightcurve_all}; namely within MJD 56160-56350 (21 August 2012- 27 February 2013) and MJD 59000-59350 (31 May 2020- 16 May 2021). The results are shown in Fig. \ref{X:infl}. The linear-Pearson correlation test applied to the data during the first flare (MJD 56160-56350; 21 August 2012- 27 February 2013) yields $-0.45$, the $p$-value being $1.6\times10^{-6}$ for $N=102$ observations, implying a negative correlation between the flux and photon index, i.e., when the source gets brighter, the photon index decreases (hardens). This behaviour has already been observed for many flaring blazars \citep[e.g.][]{1990ApJ...356..432G}. On the other hand, for the second flare the linear-Pearson test results in $0.47$ with a $p$-value of $0.001$ for $N=45$. This implies that during the X-ray flare the photon index softens, so a softer-when-brighter trend is observed. This shows that two major flares observed in the X-ray band for \source\ are different by their nature and are caused by different processes. Similar behavior of the X-ray flux of \source\ was already seen in the previous studies \citep[e.g.,][]{2016ApJ...816...53W, 2020ApJ...900..137W, 2021MNRAS.tmp.2396D}. 
\subsection{\textit{Swift} UVOT}
UVOT observed \source\ in all six filters, V (500-600 nm), B (380-500 nm), U (300- 400 nm), W1 (220-400 nm), M2 (200-280 nm) and W2 (180–260 nm) simultaneously with the XRT. 
All single observations of \source\ were downloaded and reduced using HEAsoft version 6.27 with the latest release of HEASARC CALDB. The data are reduced using standard procedures, by selecting source counts from a circular region of $5''$ around the source, while the background counts were estimated from a $20''$ region away from the source.
Host galaxy contributions were subtracted following \citet{2013MNRAS.436.1530R} and \citet{2010A&A...524A..43R} by assuming a flux density of $2.89$, $1.30$, $0.36$, $0.026$, $0.020$, and $0.017$ mJy for the host galaxy in the V, B, U, W1, M2, W2 bands, respectively. For the considered source extraction radius, the host galaxy contribution is $\sim50\%$ of the total galaxy flux, which is removed. {\it uvotsource} tool was used to derive the magnitudes which were converted to fluxes using the conversion factors provided by \citet{2008MNRAS.383..627P} which then were corrected for extinction using the reddening coefficient $E(B - V)$ from the Infrared Science Archive \footnote{http://irsa.ipac.caltech.edu/applications/DUST/}.

The optical/UV flux evolution in time is shown in Fig. \ref{lightcurve_all} panel e) and f) separating the flux in V, B, U and W1, M2 and W2 filters. The source flux is relatively constant at the level of a few times $10^{-11}\:{\rm erg\:cm^{-2}\:s^{-1}}$ up to MJD $56500$ (27 July 2013). A flaring activity occurred around MJD 56617-56622 (21-26 November 2013) when the flux in all filters exceeded $10^{-10}\:{\rm erg\:cm^{-2}\:s^{-1}}$. The major flaring activity started on MJD 59072 (11 August 2020) and the baseline flux level was above $10^{-10}\:{\rm erg\:cm^{-2}\:s^{-1}}$. The maximum flux of $(5.80\pm0.14)\times10^{-10}\:{\rm erg\:cm^{-2}\:s^{-1}}$ was observed in the V band on MJD $59249.26$ (04 February 2021). The maximum flux in B, U and M2 filters was also above $5\times10^{-10}\:{\rm erg\:cm^{-2}\:s^{-1}}$ while that in W1 and W2 was around $4\times10^{-10}\:{\rm erg\:cm^{-2}\:s^{-1}}$.

\subsection{Archival data}
To achieve as much as possible a complete view of the broad-band emission from \source\, we have also considered all the available multi-frequency archival measurements alongside with the data from \fermi, \textit{Swift} XRT and UVOT. 
These include a) optical data monitoring from the ASAS-SN Sky Patrol web site \footnote{https://asas-sn.osu.edu/} \citep{2017PASP..129j4502K}, b) \textit{NuSTAR} data from the observations of \source\ on 11 December 2012 (MJD 56272), 14 September 2019 (MJD 58740) and 11 October 2020 (MHD 59133) from Middei et al. 2021, submitted, and c) any other multi-frequency measurements available from the VOU\_BLazars tool \citep{voublazars} and the ASI Space Science Data Center  (SSDC) archive \footnote{https://www.ssdc.asi.it}. In addition we also considered the observations of \source\, carried out by VERITAS on June 11, 2011 \citep[MJD 55740 ][]{ 2013ApJ...762...92A} and on October 5, 2016 \citep[MJD 57697 ][]{2018ApJ...856...95A} and by MAGIC between June 15 and 28, 2015 \citep[MJD 57188-57201 ][]{2019A&A...623A.175M}. The combination of all these data sets results in an unprecedented amount of observations of \source\ covering the spectrum from radio frequencies to HE and VHE \gray\ bands over a period of nearly 13 years, from 2008 August to 2021 March. 

\section{Modeling the SEDs}\label{modeling}
In this section we use the data assembled as described above to investigate the evolution of the broadband spectrum of \source\ between 2008 August and 2021 March. 
To this end we have generated a large number of quasi contemporaneous SEDs by plotting the computed \gray\ spectra together with the data available in all other energy bands 
in each of the Bayesian intervals defined in Sec. \ref{datanal}.
To illustrate the temporal evolution of the broad band emission from \source\ in a visually effective way we have combined these SEDs to form an animation that is available as Supplementary data and at the following link: \href{https://www.youtube.com/watch?v=KkI1d4gK_UU}{\nolinkurl{youtube.com/L1yT105UGYM}}. 
Flux changes in the optical/UV, X-ray and \gray\ bands are evident. 
In the \gray\ band, the spectrum hardens together with the flux amplification, resulting in a shift of the peak of the second  component to higher energy values. During the brightest X-ray state (on MJD 59128.18; 06 October 2020), the low-energy SED component extended to the X-ray band as a consequence of a significant change of the location of the synchrotron peak from the usual $\sim10^{14}$ Hz to $\sim10^{16}$ Hz, well into the HBL regime \citep{Padovani1995,Abdo_2010}. Such a large modification, never observed before in \source\,, marks the extraordinary nature of this flare, which has been studied also by \citet{2021MNRAS.tmp.2396D} and  \citet{2021MNRAS.507.5602P}. 

The  classical double-humped SED of \source\ is usually interpreted within leptonic scenarios. The EGRET observations of \source\ \citep{1999ApJ...515..140S, 1999ApJ...521..145M} revealed that modeling of the HE data requires a component that extends beyond the SSC radiation generated in a single emission zone: one-zone leptonic modeling requires a very high Doppler factor ($\delta$; $\delta \simeq\Gamma_{\rm jet}$, where $\Gamma_{\rm jet}$ is the bulk Lorentz factor) or an extended emission region. Since then the SED of \source\ has been conventionally modelled within two-zone scenarios \citep[e.g.][]{2011ApJ...730..101A, 2019A&A...623A.175M} or assuming inverse Compton scattering of 
external (EIC) photons \citep[e.g., ][]{1999ApJ...521..145M, 2013ApJ...768...54B}. External Compton scenarios are favoured in \source, considering the detection, although weak, of the $H\alpha$ line \citep{1996MNRAS.281..737C,2010A&A...516A..59C}, which points to the presence of a broad-line region (BLR). Even if this BLR is not large enough to absorb VHE \grays\ through $\gamma-\gamma$ interaction \citep[e.g., ][]{2003APh....18..377D}, it can provide targets for inverse Compton up-scattering. For example, by modeling the  SED of \source, \citet{2011ApJ...730..101A} showed that the SSC+ERC scenario provides reasonable modeling of the data also during the low state and the inverse Compton scattering of the BLR-reprocessed radiation strongly dominates over that directly from the disc.

In an effort to understand the processes dominating in the jet in different physical conditions we investigated the broadband emission from \source\, by modeling the SEDs observed in different periods. 
From the SED/light curve animation discussed above we have selected all periods with sufficient multiwavelength data, typically those with flux measurements in at least the optical/UV, X-ray and \gray\ bands. This allowed us to assemble 511 high-quality and quasi-simultaneous SEDs representing \source\ in a variety of emission states.
All these SEDs are modeled assuming that the emission region ('the blob') is a sphere with radius $R$, including a magnetic field of intensity $B$ and a population of relativistic electrons following an energy distribution defined by a power law with an exponential cutoff, as expected from shock accelerations:
\begin{equation}
N(\gamma')=N'_e\gamma'^{-p}\: Exp(-\gamma'/\gamma'_{\rm cut})
\end{equation}
for $\gamma'>\gamma'_{\rm min}$ where $\gamma'_{\rm min}$ and $\gamma'_{\rm cut}$ are the minimum and cut-off energy of the electrons, respectively. It is assumed that the emission region is located inside the BLR and the low energy SED component is interpreted as synchrotron emission of relativistic electrons, while the second SED component is due to inverse Compton up-scattering of photon fields from the jet itself \citep[SSC model e.g.,][]{maraschi, bloom} and those reprocessed from the BLR clouds \citep[EIC BLR;][]{sikora}. The BLR is assumed to be spherical shell with an average radius of $R_{\rm BLR}=7\times10^{16}$ cm and lower and upper boundaries of $0.9\times R_{\rm BLR}$ and $1.2\times R_{\rm BLR}$, respectively \citep{2003APh....18..377D}. The BLR reflects 10\% of the disc luminosity whose emission is approximated as a mono-temperature black body with a luminosity of $L_{\rm d}=3\times10^{43}\:{\rm erg\: s^{-1}}$. This luminosity was estimated with a requirement that the disc component does not overproduce the optical/UV data in any period. We note that $R_{\rm BLR}$ and $L_{\rm d}$ define the density of the external photon fields, so their small changes do not affect the results and only will result in moderate changes in the normalization of the electrons.\\
\indent Our 511 SEDs represent an ample variety of different states of \source\ and in some periods the simple one-zone model described above cannot explain the observed data. For example, when the X-ray spectrum softens neither the synchrotron component (defined by the optical/UV data) nor SSC component which has a rising shape cannot account for the X-ray flux. In these cases the SEDs were modeled considering  two-zone scenarios, assuming that one region is within the BLR and the other is outside \citep[e.g., see Fig. 2 panel c) in ][]{2011A&A...534A..86T}.\\
\indent The broadband SEDs model fitting was carried out using the open source package {\it JetSet} \citep{2006A&A...448..861M, 2009A&A...501..879T, 2011ApJ...739...66T,2020ascl.soft09001T}. The free model parameters ($p$, $\gamma_{\rm cut}$, $\gamma_{\rm min}$, $\delta$, $R$ and $B$) are constrained by using the Minuit optimizer and then improved by Markov Chain Monte Carlo (MCMC) sampling of their distributions. The quality of the fits was checked by calculating the goodness-of-fit and by checking MCMC diagnostic plots. In principle, $R$ can be constrained either from the variability consideration or from SED fitting. If high quality data are available, detailed variability studies may constrain the radius from the relation $R\leq \delta\:t\: c/(1+z)$. However, in the current case, due to the high number of the considered periods for the modeling (511),  proper variability studies in each period are impossible. Therefore, in the SED fitting $R$ is considered as a free parameter allowing to vary within the range defined by the applied model, i.e., the emission region is inside the BLR. 
When two-zone modeling was considered, to reduce the number of free parameters, different but fixed radii were used for the emitting regions.
 Since the TeV or X-ray observations in the bright states reveal that the flux varies on minute scales, implying that the emitting region outside BLR should be very compact, $R=10^{15}$ cm was assumed. On the other hand, the optical/UV and \gray\ fluxes vary albeit not on such short scales, so for the blob within the BLR $R=10^{16}$ cm was used. Also, both emitting regions were assumed to have the same Doppler boosting factor. In principle, because of orientations those regions can have different Doppler boosting factors which, however, will introduce an additional free parameter.
\subsection{SEDs modeling results}
\begin{figure*}
	\includegraphics[width=0.48\textwidth]{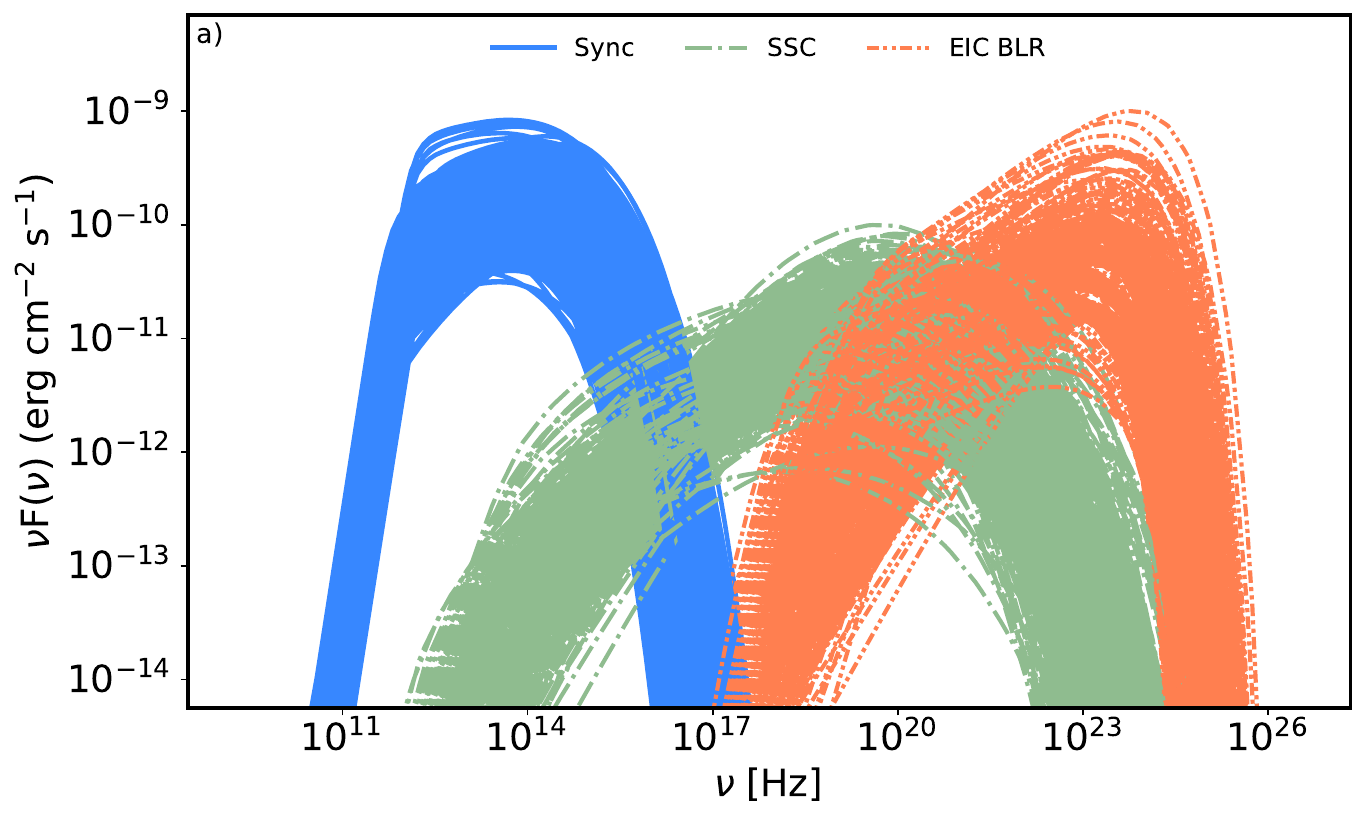}
	\includegraphics[width=0.48\textwidth]{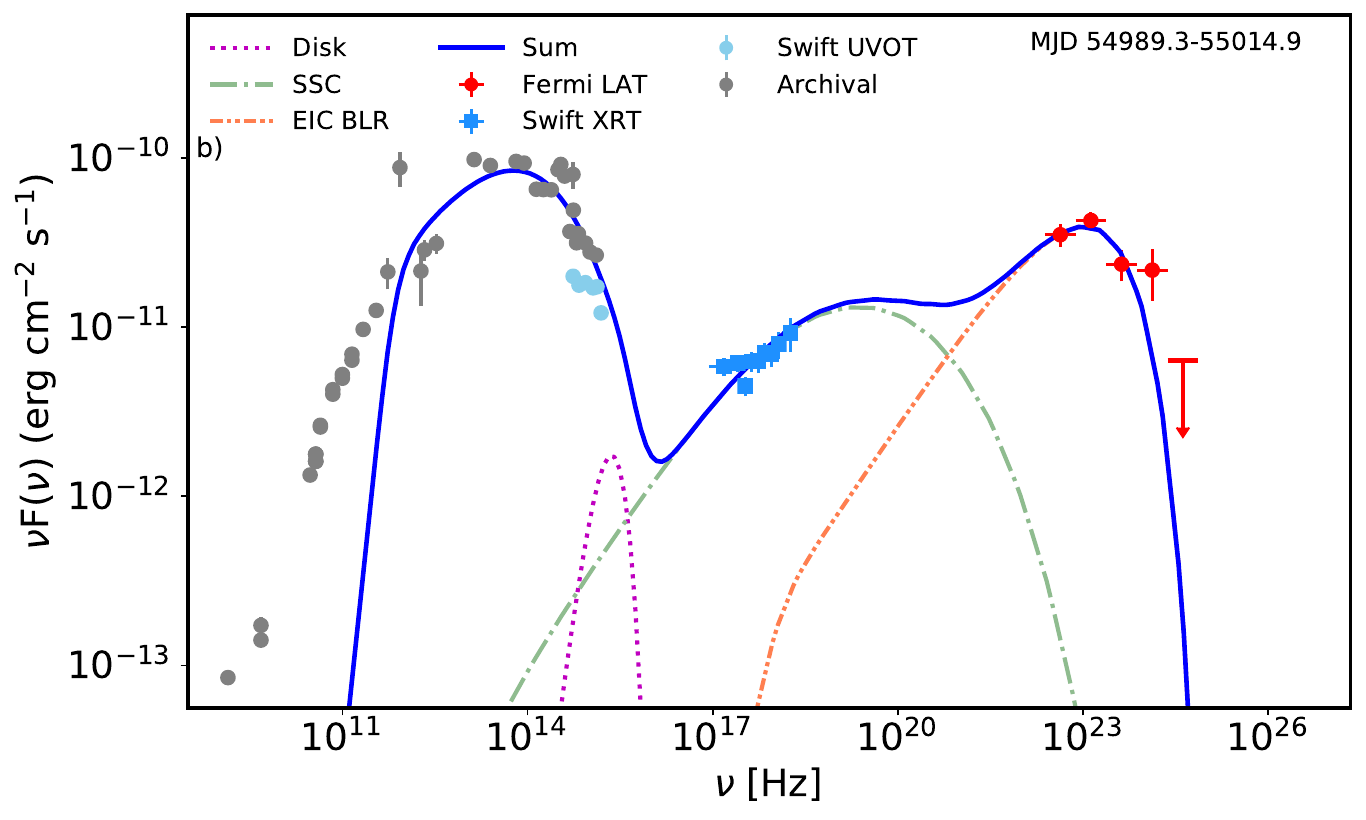}\\
	\includegraphics[width=0.48\textwidth]{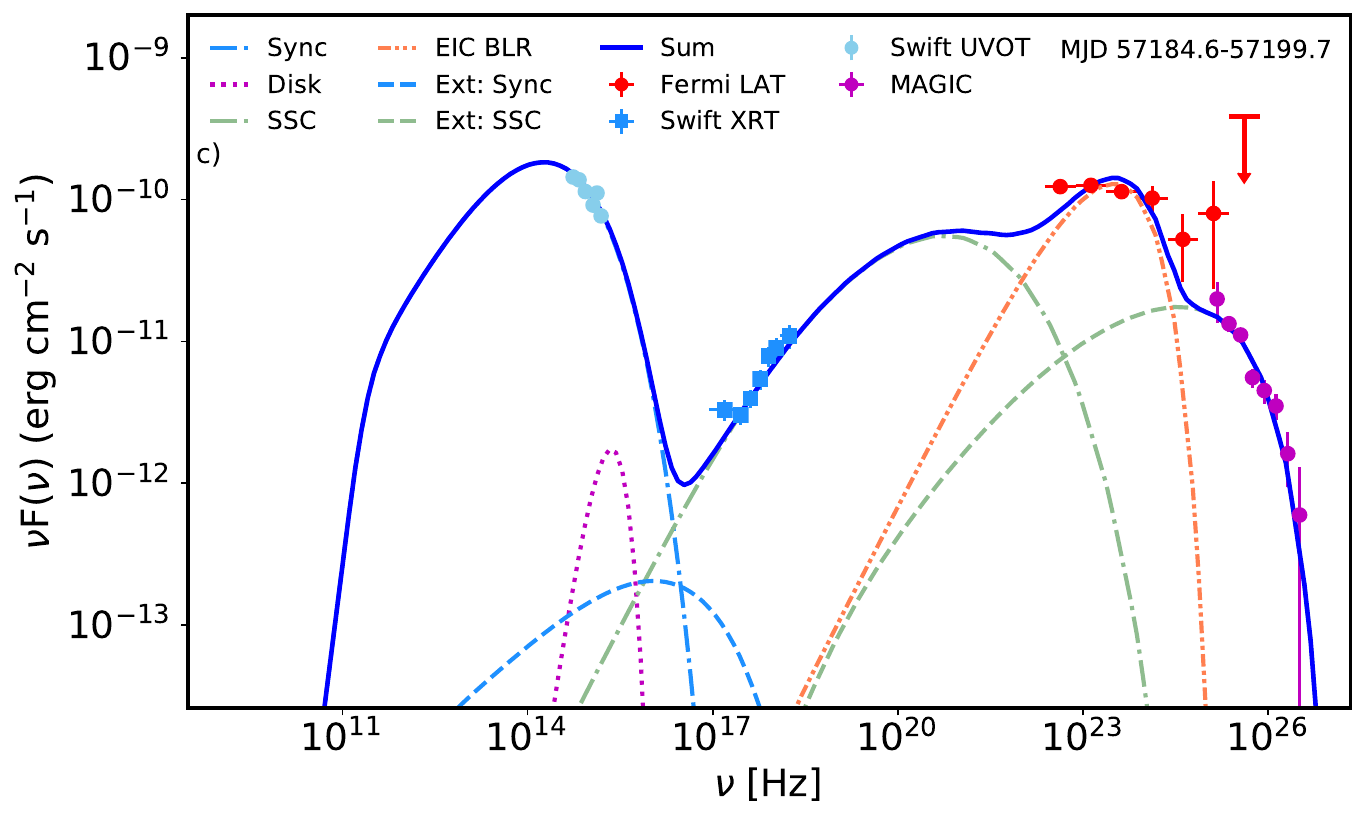}
	\includegraphics[width=0.48\textwidth]{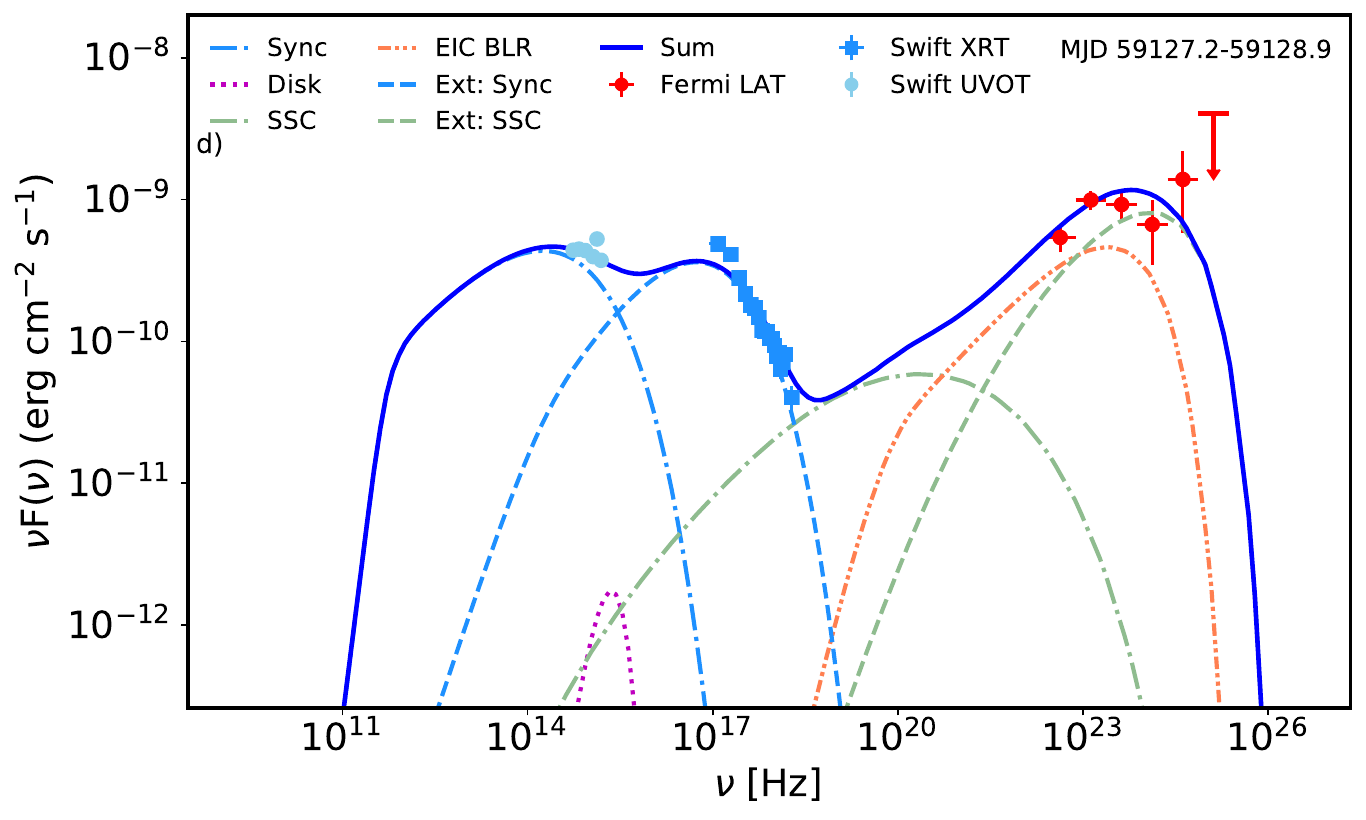}
    \caption{The broadband SEDs of \source\ observed in different emission states. {\it a)} The emission components obtained from the modeling of all SEDs with contemporaneous data collected during 2008 August-2021 August. {\it b)} The SED of \source\ in the normal emission state. {\it c)} \source\ SED modeling with VHE \gray\ data from MAGIC observations \citep{2019A&A...623A.175M}; corrected for extragalactic background light absorption using the model of \citet{2011MNRAS.410.2556D}. {\it d)} The SED of \source\ during the brightest X-ray period modeled within two-zone scenario.}
    \label{seds}
\end{figure*}

\indent {The animation of our 511 high-quality and quasi-contemporaneous SEDs of \source\ together with the corresponding modeling is available as Supplementary data and at the following link  \href{https://www.youtube.com/watch?v=J3gOi4STZCA}{\nolinkurl{youtube.com/watch?v=f3a5CGukbbE}}. In this animation, the sum of all model components is plotted as a solid blue line, while the SSC and EIC components appear as green and orange lines, respectively. The disc emission, approximated as a black body with intensity that is always below the synchrotron component, is shown in magenta. Fig. \ref{seds} shows the emission components in all 511 SEDs (panel a) and some frames representing special states (panels b-d). 
The optical/UV data constrain the tail of the synchrotron component which peaks at $\sim10^{14}$ Hz and, despite large flux variability, it remains almost unchanged as can be seen from Fig. \ref{seds} panel a) (blue curves; the bright and soft X-ray periods were not considered). The SSC emission of the synchrotron emitting electrons starts to dominate around $10^{17}$ Hz extending up to $\sim10^{23}$ Hz (green dot-dashed lines in Fig. \ref{seds} panel a) while at higher frequencies EIC of BLR photons dominates (orange dot-dot-dashed lines in Fig. \ref{seds} panel a). The change of intensity of these components show high-amplitude variability of \source\ emission in the optical/UV, X-ray and \gray\ bands. The variability in the radio band cannot be tested, as the data are missing for most of the cases. Moreover, the radio emission at lower frequencies can be produced, with significant time-lags, by the low-energy electrons in extended regions which is not associated with the emission in other bands.}
\begin{figure*}
	\includegraphics[width=0.98\textwidth]{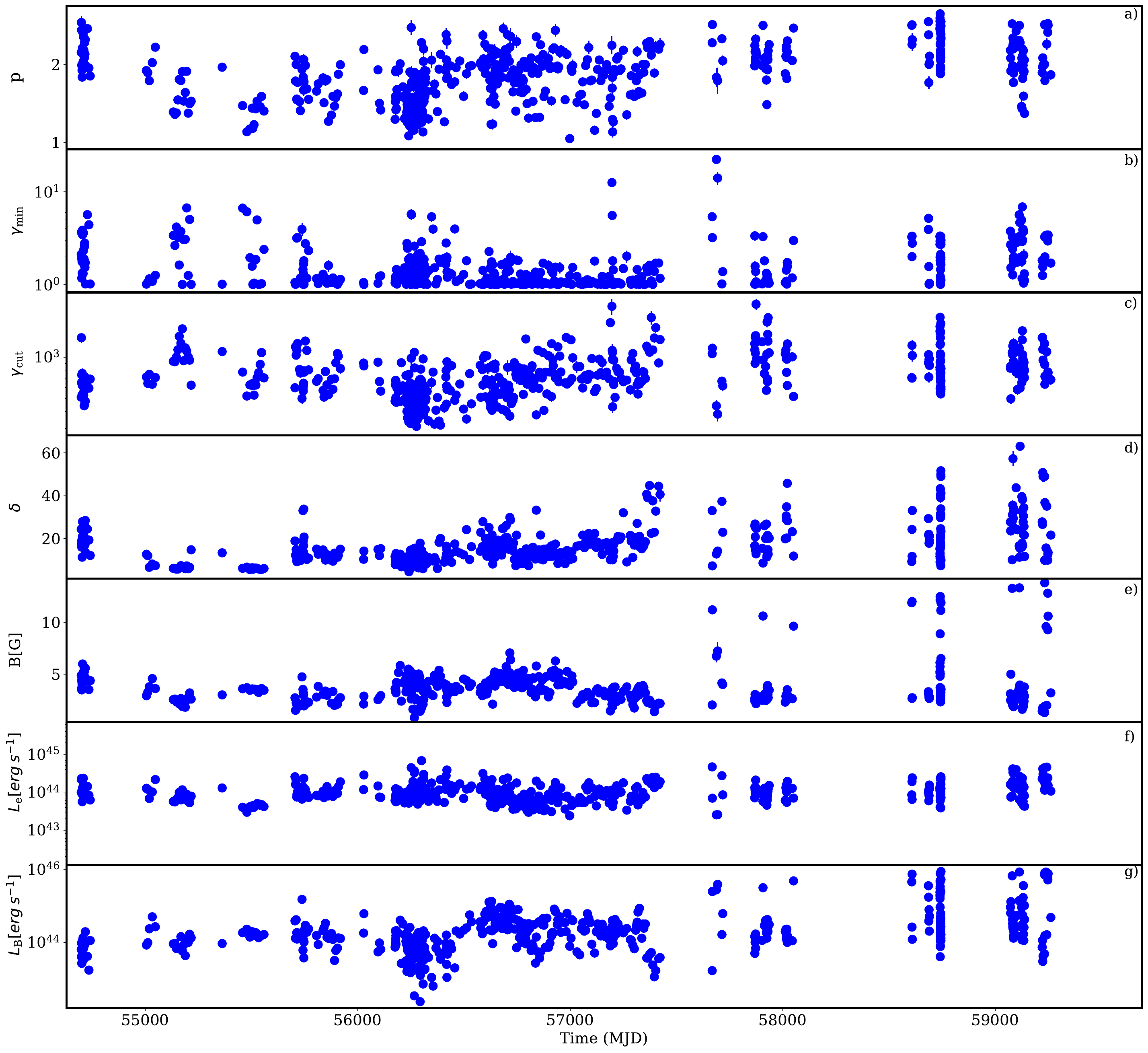}
    \caption{The evolution of electron parameters obtained from the modeling of 511 quasi-contemporaneous broadband SEDs of \source. a) the power-law index of the emitting electrons, b) and c) minimum and cut-off energy of the emitting electrons, respectively. d) and e) Doppler factor and magnetic field estimated in different periods. f) and g) the jet luminosity in particles and in magnetic field.}
    \label{elecparams}
\end{figure*}

The modeling provides estimates of the physical parameters describing the emission from \source\ and allows us to investigate their changes in time. The evolution of the $p$, $\gamma_{\rm min}$, $\gamma_{\rm cut}$, $\delta$ and $B$ parameters is shown in Fig. \ref{elecparams} panels a) to e). The minimum energy of electrons $\gamma_{\rm min}$ is mostly below $10$ (Fig. \ref{elecparams} panel b) implying that even lower-energy electrons are efficiently accelerated. The power-law index ($p$) is mostly within $1.2-2.3$ (Fig. \ref{elecparams} panel a) and is defined by fitting the X-ray data with the SSC component. Its variation is 
in accordance with the changes of the X-ray photon index shown in Fig. \ref{lightcurve_all} (panel d).
The the cut-off energy obtained from the modeling of SEDs in different periods  is shown in Fig. \ref{elecparams} panel c) which is defined mostly by the optical/UV and sometimes by \gray\ data; the minimal and maximal values of the cut-off energy are $311\pm 13$ and $2438\pm 208$, respectively. Despite such change in the cut-off energy, the low- and high- energy peaks in the SED do not deviate to higher frequencies; the peak positions are determined by the product of $\gamma_{\rm cut}$ and $B$. In this case, the magnetic field (see Fig. \ref{elecparams} panel e) varies in the range of $B=0.80-14.65$ G but, in particular, when a high $\gamma_{\rm cut}$ was estimated, $B$ was around its lower level (see Fig. \ref{elecparams}). It is also interesting to investigate the evolution of the Doppler factor $\delta$ (Fig. \ref{elecparams} panel d) which was estimated under the assumption of a constant viewing angle. This parameter remained mostly below 20 but it reached the maximum value of $\delta=63$ on MJD 59117.4 (25 september 2020) when the major \gray\ flares were observed.

An example of SED modeling during the period of MJD 54989.3-55014.9 (07 June- 02 July 2009), when the source was not flaring in any band, is shown in Fig. \ref{seds} panel b). The synchrotron emission of the accelerated electrons explains the archival (gray) and optical/UV (light blue) data, while the X-ray emission (blue) is due to SSC emission. This SSC component extends up to $10^{20}$ Hz and the emission in the \gray\ band is dominated by the IC scattering of BLR photons. This shows that even in the quiescent state of the source, the external photon field (BLR) is necessary to explain HE \gray\ data. The power-law index of the emitting electrons is $p=1.92\pm0.03$ and $\gamma_{\rm cut}=715\pm35$, while the emitting regions size is $R=4.08\times10^{15}$ cm which moves with a Doppler factor of $\delta=12.7$. The system is close to the equipartition $U_{\rm e}/U_{\rm B}\simeq1.5$ with $B=2.91\pm0.12$ G.

An example of the \source\ SED modeling with VHE \gray\ data from MAGIC observations is shown in Fig. \ref{seds} panel c). When considering a one-zone scenario, the optical/UV data with a decreasing trend 
constrains the cut-off energy of the emitting electrons and the IC scattering of $1-5$ eV synchrotron or BLR photons on them will only reach the MeV/GeV band, unable to account for the VHE \gray\ data. These periods (8 among the considered SEDs) are modeled within two-zone scenarios, considering the emitting regions inside and outside the BLR.
The emission observed in the radio to HE \gray\ bands is dominated by that from the region within the BLR, and the VHE \gray\ data are explained by the SSC emission from the compact region. In the extended region ($R=10^{16}$ cm), $B=1.87\pm0.08$ G and the emitting electrons have $\gamma'^{-1.28\pm0.02}$ distribution with $\gamma'_{\rm cut}=1082\pm36$. Instead, in the compact region, requiring that its synchrotron emission is lower than that from the other region, $B=3.32\times10^{-2}$ G is estimated, implying that the electrons can be accelerated to higher energies, i.e., $\gamma'_{\rm cut}=1.39\times10^{5}$ was estimated in this case. The contribution of these electrons with $p=2.17$ starts to dominate above $\sim 30$ GeV when the spectrum of IC scattering of BLR photons decreases, explaining the data observed by MAGIC.

Similar two-zone models are also required when the soft component in the X-ray band is observed. The synchrotron component defined by the available optical/UV data effectively extends up to $10^{17}$ Hz ($\sim400$ eV), unable to account for the observed X-ray data. Thus, in the X-ray band an additional component is dominating. Among the selected periods, 38 SEDs with a soft X-ray spectrum were modeled within the two-zone scenario; an example of SED modeling when the brightest X-ray emission was observed is shown in Fig. \ref{seds} panel d). The Doppler boosting factor of both emitting regions is $\delta=16.44$ but they are filled with different distributions of electrons. For example, the X-ray emitting electrons (the region outside BLR) have $\gamma^{-1.49}$ distribution above $\gamma_{\rm min}=761$ with a cut-off energy of $1.10\times10^4$. Instead, the electrons in the other region have a softer distribution with $2.02\pm0.25$ and are accelerated only up to moderate energies of $\gamma_{\rm cut}=1559\pm213$. The magnetic field in the region outside the BLR is stronger ($B=7.28$ G) than that in the other region ($B=1.71\pm0.02$ G) which is because \text{i)} the first region has a smaller radius and \text{ii)} the synchrotron emission should be at higher frequencies, reaching up to the X-ray band.

\section{Discussion}\label{discussion}
We performed a comprehensive investigation of the large and complex luminosity and spectral variability of \source\ using the data taken by \fermi, \textit{Swift}-XRT and \textit{Swift}-UVOT between 2008 August and 2021 March. Using the unprecedented amount of the available multiwavelength data we performed an in-depth study of the origin of nonthermal emission from \source.

In all the energy bands considered, the source shows multiple periods when the flux exceeds its average level by substantial amounts. The adaptively binned light curve, encapsulating more information, provides a detailed timing view of the \gray\ flares. The maximum \gray\ flux of $F_{\rm \gamma}(>196.7\: {\rm MeV})=(4.39\pm1.01)\times10^{-6}\:{\rm photon\:cm^{-2}\:s^{-1}}$
was observed on MJD 59231.34 (17 January 2021),  associated with a flat spectral slope with photon index of $2.03\pm0.21$. This implies an energy flux, ${\rm \epsilon F_{\epsilon}}$, of $9.39\times10^{-9}\:{\rm erg\:cm^{-2}\:s^{-1}}$ in the 0.1-300 GeV energy range, 
corresponding to an isotropic \gray\ luminosity of $L_{\rm \gamma}=4\:\pi\:d_{\rm L}^2\epsilon F_{\epsilon}=(1.06\pm0.24)\pm10^{47}\:{\rm erg\:s^{-1}}$ for a $\sim307$ Mpc distance. Assuming a Doppler factor of $\delta=20$ this corresponds to $L_{\rm \gamma}/\delta^2\simeq2\times10^{44}\:{\rm erg\:s^{-1}}$ in the proper frame of the jet. This value largely exceeds the disc luminosity estimated under any reasonable assumption (e.g., $3\times10^{43}\;{\rm erg\:s^{-1}}$ to not overproduce the observed optical/UV data) implying extreme energetics 
during the \gray\ flares \citep[e.g., ][]{2014Natur.515..376G}. Unlike the flux, the photon index does not usually vary significantly, although it occasionally hardens to values $<2.0$. Such hardening, for example, was noticed after the brightening observed on MJD 59247.4 (02 February 2021) when the \gray\ flux measured within three days was $(3.32\pm0.15)\times10^{-6}\:{\rm photon\:cm^{-2}\:s^{-1}}$; the photon index was within $1.79-1.89$ during MJD 59247.4-59256.4 (02-11 February 2021).

The \textit{Swift}-XRT observations spanning different years showed an interesting behaviour of \source. Although there can be seen flux variations in different observations, there are two major flaring activities on MJD 56268.65 (07 December 2012) and 59128.18 (06 October 2020) when the source was in an elevated state for a prolonged period. Even though during these flares the flux increased almost at the same level, the photon index was significantly different. 
During the first flare, when the 0.3-10 keV flux reached it maximum value of $(8.68\pm0.84)\times10^{-11}\:{\rm erg\:cm^{-2}\:s^{-1}}$  and the traditional harder-when-brighter trend was observed. In the X-ray band, this is a known behaviour for blazars \citep{1990ApJ...356..432G,1999ApJ...527..719Z, 2005A&A...434..385G}. 
On the other hand, during the second flare the linear-Pearson test resulted in $0.47$ showing softer-when-brighter trend. Such a pattern was also observed in the X-ray emission of OJ 287 \citep[e.g., ][]{2018MNRAS.480..407K, 2021arXiv210807255G, 2021MNRAS.504.5575K}. During this flare, when the highest X-ray flux was observed (on MJD 59128.18), the spectral index is $2.84\pm0.03$ - very different from the values normally observed in \source\ (typically $\Gamma_{\rm X}\leq2.0$). There are 36 additional occasions when the X-ray spectrum softened ($\Gamma_{\rm X} \geq2.3$; see Fig. \ref{lightcurve_all} panel d) but the exceptional softening during this flare was never observed for \source. 

The softening of the X-ray spectrum also affects the peak frequency of the synchrotron component. When the soft component is associated with a high X-ray flux (i.e., $F_{\rm X (0.3-10\:keV)} > 10^{-11}\:{\rm erg\:cm^{-2}\:s^{-1}}$) the peak of the SED low-energy component reaches frequencies of $\gsim10^{15}$ - $10^{16}$ Hz, instead of the usual $\sim10^{14}$ Hz,  temporarily placing \source\, into the domain of HBL blazars. This component is present during the period MJD 59124.73-59132.88 (02-10 October 2020) when the flux rises from  $F_{\rm X (0.3-10\:keV)}= (4.83\pm0.40)\times10^{-11}\:{\rm erg\:cm^{-2}\:s^{-1}}$ with $\Gamma_{\rm X}= 2.69\pm0.11$, and reaches $F_{\rm X (0.3-10\:keV)}= (3.24\pm0.08)\times10^{-10}\:{\rm erg\:cm^{-2}\:s^{-1}}$ ($\Gamma_{\rm X}= 2.84\pm0.03$) on MJD 59128.18 (06 October 2020). 
This component fades on MJD 59132.88 (10 October 2020) when the flux decreases to $F_{\rm X (0.3-10\:keV)}= (2.44\pm0.12)\times10^{-11}\:{\rm erg\:cm^{-2}\:s^{-1}}$ ($\Gamma_{\rm X}= 2.87\pm0.12$).

During these soft states (38 among the selected 511 periods) the source's X-ray emission is driven by a new HBL-like component, which significantly differs from the usual X-ray spectrum of BL Lac. Our modeling shows that this component may come from a separate emission zone with specific properties, like the size of the emission region, the population of electrons, etc. As an example, some of the SEDs observed during the soft X-ray emission period are shown in Fig. \ref{only_syn}.  The new soft component clearly goes beyond the synchrotron radiation constrained by the optical/UV data and is interpreted as synchrotron emission from the second region (dashed line) containing much more energetic particles. For example, the electrons should be accelerated up to $\gamma_{\rm cut} =1.76\times 10^4$ with $p=1.15$ to explain the data observed on MJD 59124.73 (02 October 2020; red line). Similar parameters obtained from the modeling of the SED on MJD 59128.18 (06 October 2020) are $\gamma_{\rm cut}=1.10\times10^4$ and $p=1.49$ (blue line) but the magnetic field is $7.28$ G, significantly higher compared to the previous case ($1.65$ G). Such a large magnetic field is required because of the increase in the X-ray flux ($\sim6.11$ times) which cannot be explained by changing $\gamma_{\rm cut}$; $p$ and $\gamma_{\rm cut}$ are also constrained by the \gray\ data. The X-ray flux variation impacts the magnetic field which decreases to $B=0.27$ G on MJD  59132.88 (10 October 2020) when the soft X-ray component was with a low state. The electrons in the emitting region are still energetic with $\gamma_{\rm cut}=5.47\times10^4$ but their contribution starts to be subdominant. In fact, the usual hard X-ray component which is interpreted as SSC radiation from the blob inside the BLR dominates already on MJD 59133.81 (11 October 2020; Fig. \ref{only_syn} right panel). This implies that either the acceleration/injection of the energetic electrons is not efficient anymore or due to the drop of the magnetic field these electrons cool down on longer time scales; for example, when $B=0.1$ G the cooling of $\gamma=5\times10^4$ electrons in the observer frame is $t_{\rm cool}=6\:\pi\:m_{e}\:c\:(1+z)/\sigma_{\rm T}\:B^{2}\:\delta\gamma=1.3\:(B/0.1\:G)^{-2}\:(\delta/15)^{-1}\:(\gamma/5\times10^4)^{-1}$ day. 

The composition of the second emitting region ($U_{\rm e}/U_{\rm B}$) changes during the periods shown in Fig. \ref{only_syn}. A slightly particle dominated region ($U_{\rm e}/U_{\rm B}\simeq12.3$) is necessary to explain the SED on MJD 59124.73 (02 October 2020) while it should be  magnetically dominated with $U_{\rm e}/U_{\rm B}\simeq0.49$ to explain the data observed on MJD 59128.18 (06 October 2020) and it is strongly particle dominated on MJD 59132.88 (10 October 2020) with $U_{\rm e}/U_{\rm B}\simeq3.1\times10^3$. This indicates that the X-ray flare was caused by the injection of new energetic particles and by sudden increase of the magnetic field. This is in agreement with the observation of the softer photon index during the bright X-ray state; due to the high magnetic field the electrons cool faster forming a soft spectrum with the increase of intensity.

The observed VHE \grays\ are also most likely produced from the second emission zone, although its composition is different. This region
still contains energetic electrons with $\gamma_{\rm cut}\simeq10^5$ but the magnetic field is low ($0.02-0.1$ G), so the region is strongly particle dominated with $U_{\rm e}/U_{\rm B}\geq10^3$. This makes the emission from these electrons significant in the VHE \gray\ band with no significant contribution at lower energies.

Our modeling shows that the overall emission from \source\ from time to time is produced from two regions separated in the jet. We note that
an excess of a new component in the X-ray band was already noticed in previous observations of \source\ in 2007-2008: the XMM Newton observations showed that the X-ray spectrum was flat/concave producing a mild soft-X-ray excess, suggesting that two components are contributing in this band \citep{2009A&A...507..769R}. This excess was interpreted by the helical jet model of \citep{1999A&A...347...30V}. Also BeppoSAX observations indicated presence of two synchrotron components in the broadband SED of \source\ \citep{2003A&A...408..479R}. The formation of the second emission region can be explained in the framework of other hypotheses as well. For example, it can be a local reconnection outflow in the “jet in a jet” scenario \citep{2009MNRAS.395L..29G, 2010MNRAS.402.1649G}. The second population of energetic electrons can be formed also when the energetic protons interact in the jet; the electrons are produced from the decay of muons and are more energetic than the initial cooled electrons. The radiative signature of these electrons initially appears at HE, which in time is shifted to lower energies.
In the scenario considered by \cite{2021ApJ...906..131M} the second emission region could be where occasionally accelerated protons radiate X-ray photons via the synchrotron mechanism and interact with them via photomeson processes producing detectable high-energy neutrinos \citep{2021arXiv210714632S}. 

The extensive modeling of SEDs presented in this paper shows that
except for the cases when \source\ was in a high and soft X-ray emission or a flaring VHE \gray\ state (46 in total), a one-zone leptonic model involving inverse Compton scattering of synchrotron and BLR-reprocessed photons gives a reasonable modeling of the data. 
The parameters values obtained from our modeling  are similar to those typically estimated for blazars; their evolution in time is shown in Fig. \ref{elecparams}. In some cases, the power-law index of the emitting electrons (panel a) on Fig. \ref{elecparams}) should be hard ($<2.0$) to account for the observed X-ray data. Such a hard injection index is difficult to obtain within standard shock accelerations, but it can be achieved when the particles are accelerated via magnetic reconnection \citep{2014ApJ...783L..21S}. For example, \citet{2014PhRvL.113o5005G} using fully kinetic simulations demonstrated that in highly magnetized environments ($\sigma>>1$) the spectral index of the particles approaches $p=1.0$; the $\sigma>>1$ condition is required so that the time scale over which particles are injected into the acceleration region is longer than the first-order Fermi acceleration time. The cut-off energy of the emitting electrons takes value between
$311\pm 13$ and $2438\pm 208$ and shows that the electrons are efficiently accelerated up to $\sim1$ GeV during the flares. The cut-off energy is naturally formed when the acceleration is limited by the cooling or dynamical time scale \citep[e.g.,][]{2013ApJ...765..122Y, 2018PASP..130h3001Z, 2020A&A...635A..25S, 2017MNRAS.464.4875B}. For example, when the particles are abruptly injected into the emitting region, they start to loose energy or escape the region, so the HE tail of the particle distribution steepens and a cut-off is formed. In principle, the cut-off energy values given in Fig. \ref{elecparams} panel c) can be obtained under a reasonable assumption for the injection and escape times. It should be noted that a similar cut-off feature in the electron spectrum will be formed also in the case of an episodic injection with an energy-dependent escape. 

The comparison of the multiwavelength light curve shown in Fig. \ref{lightcurve_all} with the Doppler boosting factor evolution in time given in Fig. \ref{elecparams} panel d) shows that it substantially increases when the source is bright in the \gray\ band. This is a consequence of the current interpretation the \gray\ data as inverse Compton scattering of the external photon field whose density transferred to the jet frame is $u_{\rm ph}^{'}\sim \delta ^2\:u$. Therefore, any increase in the \gray\ luminosity would require a larger $\delta$. Although the values estimated for \source\ and shown in Fig. \ref{elecparams} are not physically unrealistic, in other interpretations, e.g., in two zone emission scenarios, a lower value of $\delta$ would be acceptable. 

The modeling allows us to assess the luminosity of the jet in various periods. The evolution of the power (luminosity) carried by the jet in the form of electrons and magnetic field computed as $L_{e}=\pi c R_b^2 \Gamma^2 U_{e}$ and $L_{B}=\pi c R_b^2 \Gamma^2 U_{B}$ respectively, is shown in Fig. \ref{elecparams} panels f) and g), respectively. Both are relatively constant with a mean luminosity of $L_{\rm e, mean}=1.12\times10^{44}\:{\rm erg\:s^{-1}}$ and $L_{\rm B, mean}=5.09\times10^{44}\:{\rm erg\:s^{-1}}$, but they slightly increased during flaring periods. The jet is magnetically dominated in the periods when the optical/UV data (defined by the synchrotron component) exceeds the X-ray data (defined by SSC), for the other cases $L_{\rm e}/L_{\rm B}>1$ was estimated. The total luminosity ($L_{\rm tot}=L_{\rm e}+L_{\rm B}$) varies within $(0.07-8.86)\times10^{45}\:{\rm erg\:s^{-1}}$ which is lower than the Eddington luminosity $4.75\times10^{46}\:{\rm erg\:s^{-1}}$ for a black hole mass of $3.8\times10^8\:{\rm M_{\rm BH}}$ 
\citep{2009RAA.....9..168W}; see also \citet{2003MNRAS.343..505F, 2010A&A...516A..59C, 2010MNRAS.402..497G, 2017A&A...602A.113T}. Moreover, this condition will be still satisfied when considering the second emission region with $L_{\rm tot}=(0.1-8.1)\times10^{44}\:{\rm erg\:s^{-1}}$, comparable to the luminosity of the other region.

The multiwavelength SED of \source\ observed in different periods has been modeled within various scenarios \citep[e.g.,][]{1999ApJ...521..145M, 1999ApJ...515..140S, 2002A&A...383..763R, 2000AJ....119..469B, 2011ApJ...730..101A, 2016ApJ...816...53W, 2019A&A...623A.175M}. For example, in \citet{2011ApJ...730..101A} the synchrotron/SSC, two-zone SSC and synchrotron/SSC plus EIC models were considered to fit the averaged (2008 August 20–September 9) SED of \source. The SSC and EIC emission of electrons initially injected with a $2.85$ power-law index in the emitting region with a radius of $3\times10^{15}$ cm which moves with a bulk Doppler factor of 15 can reproduce the observed data. This model is preferred also from the viewpoint of equipartition considerations, i.e., $L_{\rm B}/L_{\rm e}=1.48$. Alternatively, \citet{2019A&A...623A.175M} considered a two-zone scenario for modeling the VHE \gray\ flare of \source, discussing a different setup for the emitting regions. Assuming a smoothed broken power-law distribution for the emitting electrons, correspondingly $2.0$ and $3.2$ ($3.7$) indices were estimated before and after the break for the compact (extended) emitting region. The minimum energy of the electrons is $50$ and $3.0$ for the compact ($10^{15}$ cm) and extended ($10^{17}$ cm) emitting regions, respectively, which move with a Doppler factor of $60$ and $7$, respectively. The parameters obtained here are not substantially different from those usually estimated for \source. The long-term flux variability of \source\ is also discussed in the context of the geometrical changes, i.e., in the jet the emitting regions have different orientations with respect to the line of sight \citep{2013MNRAS.436.1530R}.

\section{Conclusion}\label{concl}
In this paper we have presented a long-term (thirteen-year-long) multiwavelength study of the peculiar blazar \source. Using an adaptive binning method for the generation of the \gray\ light curve a very different state of the source emission was identified and studied, revealing complex and high-amplitude variability. Thanks to the good X-ray coverage 
(610 \textit{Swift} XRT observations), two major X-ray flaring activities were identified. Although X-ray flux variations are common in \source, the observed flaring activities showed substantially different properties; during the flare observed on MJD 59128.18 (06 October 2020) the flux increase was associated to a X-ray photon index softening to $2.84\pm0.03$, resulting from the shift of the synchrotron peak to higher frequencies. We investigated the evolution of the X-ray photon index in time and identified additional 38 periods when the X-ray photon index softens, extending the X-ray emission beyond the synchrotron component extrapolated from the optical/UV band.

We also performed a comprehensive modeling of \source\ SEDs selected in different periods. Most of the time the broad-band emission of the source can be described within a simple one-zone scenario when the emission region is inside the BLR, considering the inverse-Compton up-scattering of both synchrotron and BLR reprocessed photons. However, in the periods when the X-ray emission is associated to a soft spectral index and when VHE \grays\ were observed, the data could be modeled only considering a second emitting region outside the BLR. The modeling shows that, depending on the magnetic field and the $U_{\rm e}/U_{\rm B}$ ratio, the radiative signature of the second emitting region contributes to either the X-ray or VHE \gray\ bands. The model parameters estimated through fitting 511 broadband SEDs allow us to track the changes in the jet that are responsible for multiwavelength flares.

The accumulation of a large number of high-quality data from the observations in different bands provides an exceptional chance to investigate the dynamical evolution of jet radiation in time. Through this new comprehensive approach the main properties defining the jet physics can be compared and contrasted, helping to unveil the origin of the emission in different periods. 

\section*{Acknowledgements}
We acknowledge the use of data, analysis tools and services from the Open Universe platform, the ASI Space Science Data Center (SSDC), the Astrophysics Science Archive Research Center (HEASARC), the Fermi Science Tools, the All-Sky Automated Survey for Supernovae (ASAS-SN), the Astrophysics Data System (ADS), and the National Extra-galactic Database (NED).

This work was supported by the Science Committee of RA, in the frames of the research project No 20TTCG-1C015.

This work used resources from the ASNET cloud.
\section*{Data availability}
All data used in this paper is public and is available from the \textit{Swift}, Fermi and \textit{NuSTAR} archives, from the Open Universe tools and on-line services or from the ASAS-SN web page.
The SED data used for model fitting can be shared on reasonable request to the corresponding author.

\section*{Supporting Information}
Supplementary data are available at MNRAS online.



\bibliographystyle{mnras}
\bibliography{biblio} 








\bsp	
\label{lastpage}
\end{document}